\documentclass[12pt,a4paper,english,pra,titlepage,showpacs,preprint,showkeys]{revtex4}
\usepackage[latin1]{inputenc}
\usepackage{amsmath}
\usepackage{amssymb}
\usepackage{graphicx}
\usepackage{epstopdf}
\usepackage{babel}

\makeatletter
\@ifundefined{textcolor}{}
{
 \definecolor{BLACK}{gray}{0}
 \definecolor{WHITE}{gray}{1}
 \definecolor{RED}{rgb}{1,0,0}
 \definecolor{GREEN}{rgb}{0,1,0}
 \definecolor{BLUE}{rgb}{0,0,1}
 \definecolor{CYAN}{cmyk}{1,0,0,0}
 \definecolor{MAGENTA}{cmyk}{0,1,0,0}
 \definecolor{YELLOW}{cmyk}{0,0,1,0}
}
\makeatother

\begin{document}

\title{Quantum erasure in the presence of a thermal bath: the effects of system-environment microscopic correlations}
\author{A. R. Bosco de Magalhães}
\email{magalhaes@des.cefetmg.br}
\author{J. G. Peixoto de Faria}
\email{jgpfaria@des.cefetmg.br}
\affiliation{Departamento de Física e Matemática\\
Centro Federal de Educação Tecnológica de Minas Gerais\\
Av. Amazonas 7675, Nova Gameleira\\
Belo Horizonte, MG, CEP 30510, Brazil}
\author{R. Rossi Jr.}
\email{romeu.rossi@ufv.br}
\affiliation{Universidade Federal de Viçosa --- Campus Florestal\\
Florestal, MG, CEP 35690-000, Brazil}

\begin{abstract}
We investigate the role of the environment in a quantum erasure setup in the
cavity quantum electrodynamics domain. Two slightly different schemes are
analyzed. We show that the effects of the environment vary when a scheme is
exchanged for another. This can be used to estimate the macroscopic
parameters related to the system-environment microscopic correlations.
\end{abstract}

\maketitle

\section{Introduction}

\textquotedblleft In reality, it contains the only mystery, the basic
peculiarities of all of quantum mechanics.\textquotedblright\ \cite{art1}.
The famous statement by Richard Feynman about the wave-particle duality,
gives a glance on the relevance of the subject to quantum theory. The double
slit experiment illustrates very well the wave-particle duality. In such
an experiment, if the information about which slit the quanton (in the sense
of \cite{art2,art3}) has crossed (which-way information) is available, the interference fringes are not visible on the screen (particle behavior); however,
if the which-way information is not available, there is an interference
pattern (wave behavior).

At the Solvay conference (1927), A. Einstein presented a \textit{gedanken} experiment
(the Recoiling Slit Experiment) which consisted in a double slit experiment
with a movable slit placed before the double slit. The goal was to detect
which-way information of the quanton (recorded by the movable slit) and still
see an interference pattern \cite{art4}. The apparent difficulty imposed by
such \textit{gedanken} experiment was solved by N. Bohr, who pointed out that a
careful analysis of the movable slit would require the inclusion of
uncertainty relations of its position and momentum; this would add random
phases in the quanton path and consequently it would make the interference
pattern vanish. Therefore, in this argumentation, N. Bohr used the
uncertainty principle to sustain the wave particle duality. 

In the eighties, another \textit{gedanken} experiment, the quantum eraser, proposed by M.
Scully and collaborators \cite{art5,art6,art7}, brought back to light the
debate about wave-particle duality. In the quantum eraser experiments, the
quanton interacts with a probe system, and they become entangled. This
interaction makes the which-way information available and destroys the
interference pattern, even when there is no relevant modifications on the
quanton position and momentum degrees of freedom. According to the authors,
the entanglement is the essential key behind this phenomenon, and it is not
necessary to call upon Heisenberg's uncertainty principle, as it was done in
the early discussions between A. Einstein and N. Bohr. As a result, a debate on the role
of the entanglement and uncertainty relations began \cite%
{art8,art9,art10,art11,art12}. In a quantum eraser experiment, the which-way
information available in the entangled state can be erased, and consequently
the interference pattern recovered, by correlating the measurement results of
the probe and the interferometric system.

Several experimental observations of the quantum eraser have been reported 
\cite{art13,art14,art15,art16,art17,art18}. The quantum eraser is an
important tool for debating on fundamental questions, but it is also used in
practical applications. To quote a few examples: In Ref. \cite{art19}, it was
used as a tool for channel corrections; in Ref. \cite{art20}, to improve
the cavity spin squeezing; in Ref. \cite{art21}, for imaging applications; in Ref. \cite{art22}
for experimental entanglement verification.

In the present work, we propose an experimental setup in the context of cavity quantum electrodynamics
(CQED) where quantum erasure can be accomplished. We propose an implementation of Ramsey
interferometry, where the interferometric paths are represented by two internal states of Rydberg
atoms. The which-way information is held by a bipartite system composed of two microwave
modes that interact with a common environment modeled by a thermal reservoir \cite{art23,art24}. 
 
The correlation resulting from the interaction between two or more systems and a 
common bath is responsible for the appearance of a set of states that are robust against 
decoherence \cite{palma1996,duan1998}---the decoherence-free subspaces. 
In our model, the coupling of the bipartite 
system to a common bath leads to a dramatic difference in the
process of erasure depending on the class of states chosen to perform it.
This fact allows us
to propose a measurement scheme of the parameter related to the cross correlations resulting
from the interaction between the bipartite which-way register and the environment.

This work is organized as follows. In the next section, we describe the basic setup and compute the
action of the bath. In Section 3, we describe how a slightly modified scheme
can be used to highlight cross decay rates related to the cavity modes.
The Conclusion is found in Section 4.

\section{Setup for quantum erasure}

In the Ramsey interferometry experiments, the interference is observed by
dealing with the states of the internal degrees of freedom of atoms or molecules:
the role of different paths in this type of interferometry is played by
them. Accordingly, it is necessary to create coherent superpositions of
these states with the ability to manipulate the relative phase. When
counting the number of atoms or molecules in a given state as
a function of the relative phase, the interference can be observed.
Clearly, the which-way information destroys the interference, and the erasure of
this information may be used to restore it. The proposed experiment is based
on Ramsey interferometry. We examine a setup where there is which-way
information related to the states of two microwave modes. The scheme
involving two modes is more complex than that possibly designed with only one
mode; nevertheless, as will be clear, two cavity modes are necessary to
investigate system-environment microscopic correlations. 
\begin{figure}[tbp]
\includegraphics[scale=2.0]{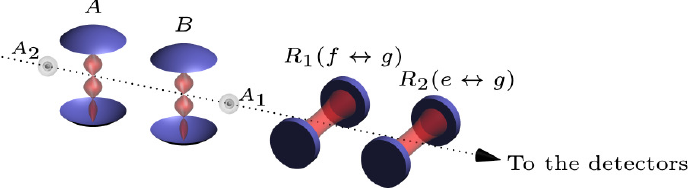}
\caption{CQED experimental setup devised to detect the quantum erasure
assisted by the environment. A Ramsey interferometer is implemented using
two microwave cavities, $R_{1}$ and $R_{2}$, where classical fields are
stored, resonant or quasi-resonant with transitions $f\leftrightarrow g$ and 
$e\leftrightarrow g$, respectively. $A$ and $B$ are high-Q microwave
cavities that work out as path identifiers. The atomic levels $e$ and $g$
play the role of interferometric paths of the atom $A_{1}$. The second atom $%
A_{2}$ is used to probe the state of the high-Q cavities $A$ and $B$, and the
erasing of the which-path information is yielded by coincidence measurements
of the states of the two atoms.}
\end{figure}

Consider two superconducting cavities $A$ and $B$ that support the resonant
modes $M_{A}$ and $M_{B}$ with frequency $\omega $. Atoms with levels $i$, $e
$, $f$, and $g$ relevant to the experiment will go through these cavities and
two Ramsey zones (see Fig. 1). The frequencies related to the atomic
transitions are illustrated in Fig. 2. The tuning of the atomic transitions
with the field modes can be performed by means of the Stark effect. The
transition $e\rightarrow g$ is assumed to be resonant with the modes $M_{A}$
and $M_{B}$ when there is no Stark effect. We assume, in what follows, that
the relations between couplings and detunings are such that non resonant
transitions can be ignored at every step.

\begin{figure}[tbp]
\includegraphics[scale=1.0]{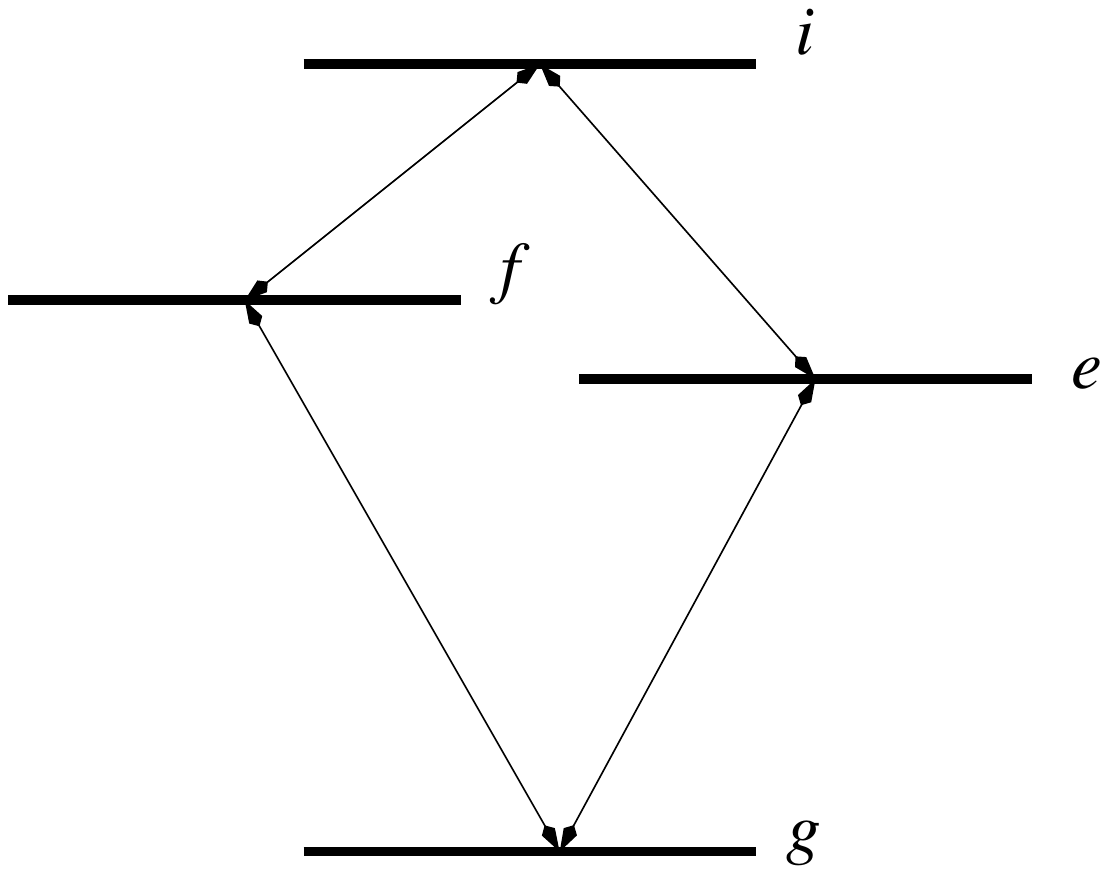}
\caption{Scheme of the relevant levels of the Rydberg atoms used in the
experiment. The transition between the $e$ and $f$ levels is not allowed.
Classical fields in one of the Ramsey zones are resonant or quasi-resonant
with the transition $f\leftrightarrow g$. The second Ramsey zone is resonant
or quasi-resonant with $e\leftrightarrow g$.}
\end{figure}

We consider the initial state of the cavities as the vacuum state. An atom prepared in
the state $i$ is sent, first passing through the cavity $A$ and then the cavity $%
B$. When the atom enters the cavity $A$, the $i\rightarrow e$ transition is
brought into resonance with the mode $M_{A}$ by the Stark effect during one $\pi
/2$ Rabi pulse. Next, it goes to the cavity $B$, where the transition $%
i\rightarrow f$ is put into resonance with the mode $M_{B}$ for one $\pi $
Rabi pulse. Then, the atom flies to a Ramsey zone tuned with the transition $%
f\rightarrow g$. After this transition is performed, the atom travels to
another Ramsey zone, tuned with the transition $e\rightarrow g$. With a
suitable choice of the atomic dipole, the state of the system just before
this Ramsey zone will be 
\begin{equation}
\left\vert \psi_1\right\rangle =\frac{\left\vert g_{1}\right\rangle
\left\vert 0_{A}\right\rangle \left\vert 1_{B}\right\rangle +e^{i\phi
_{1}}\left\vert e_{1}\right\rangle \left\vert 1_{A}\right\rangle \left\vert
0_{B}\right\rangle }{\sqrt{2}},
\end{equation}%
where $\phi_1$ depends on the energies of the modes and atomic states, as
well as the distances in the experimental apparatus and the velocity of the
atom. When the atom passes through the second Ramsey zone, the system
evolves \ to 
\begin{equation}
\left\vert \psi_2\right\rangle =\frac{1}{2}\left[ \left\vert
e_{1}\right\rangle \left( \left\vert 0_{A}\right\rangle \left\vert
1_{B}\right\rangle +e^{i\phi_1}\left\vert 1_{A}\right\rangle \left\vert
0_{B}\right\rangle \right) -\left\vert g_{1}\right\rangle \left( \left\vert
0_{A}\right\rangle \left\vert 1_{B}\right\rangle -e^{i\phi_1}\left\vert
1_{A}\right\rangle \left\vert 0_{B}\right\rangle \right) \right] .
\end{equation}%
The probability of finding the atom in the state $e$ ($g$) is given by $%
P_{e}=1/2$ ($P_{g}=1/2$), showing no interference.

In ordinary Ramsey interferometers, the aim in the first step is to create a
state of superposition between the atomic levels $e$\ and $g$\ with a
relative phase that can be varied. The preparation of the state $%
\left\vert \psi_1\right\rangle $\ is analogous to this first step.
However, this state clearly exhibits a perfect path discrimination due to the
entanglement between the atom and the cavity modes. This prevents the direct
observation of the interference between the paths related to $e$\ and $g$\ as in
usual Ramsey interferometers. To observe the interference, it is
necessary to perform the erasure of the which-way information. In the
original proposal for quantum erasure, this was performed by a detector
that interacted only with the symmetric mode of the field. Here, the erasure
is achieved by sending a second atom that absorbs only the energy of the
symmetric mode or of the antisymmetric mode of the field. As we will see,
the action of the environment can vary according to this choice.

In order to investigate how the bath can disturb the erasure process, we permit
a time interval $\tau$ between the end of the interaction of the field with
the first atom and the beginning of the interaction with the second atom.
The environment, at zero temperature, will be considered only during this
interval, which should be large compared to the other times involved in the
experiment. The decay of the fields and atomic states should be slow enough
so that they can be neglected outside this interval. Experiments with atoms
with slow decay and cavities with very high quality factors have been
reported \cite{art26}. 

For the bath at zero temperature, the action of the environment can be
computed using the master equation 
\begin{equation*}
\frac{d}{dt}\rho =\mathcal{L}\rho .
\end{equation*}%
Here, $\rho $ refers to the state of the field modes $M_{A}$ and $M_{B}$ and
the Liouvillian $\mathcal{L}$ is given by 
\begin{eqnarray}
\mathcal{L} &\cdot =&-i\left[ \omega a^{\dagger }a+\omega b^{\dagger
}b,\cdot \right]  \\
&&+k\left( 2a\cdot a^{\dagger }-\cdot a^{\dagger }a-a^{\dagger }a\cdot
\right) +k\left( 2b\cdot b^{\dagger }-\cdot b^{\dagger }b-b^{\dagger }b\cdot
\right)   \notag \\
&&+k_{c}\left( 2a\cdot b^{\dagger }+2b\cdot a-\cdot b^{\dagger }a-\cdot
a^{\dagger }b-a^{\dagger }b\cdot -b^{\dagger }a\cdot \right) ,  \notag
\end{eqnarray}%
where $a^{\dagger }$ ($b^{\dagger }$) and $a$ ($b$) are the creation and
annihilation operators related to the modes $M_{A}$ ($M_{B}$), respectively, 
$k$ is the decay rate of both the cavity modes, and $k_{c}$ is the cross decay
rate. We use the usual notation for superoperators, where the symbol $\cdot $
indicates where the density operator must be placed. According to the
discussion presented in \cite{art25}, we consider $k_{c}\geq 0$.

Let us return to the experimental sequence. Suppose that the state of the first
atom is measured immediately after it passes through the second Ramsey zone.
If the result of this measurement is $e$, the state of the field modes after
the interval $\tau$ will be 
\begin{equation*}
\rho _{M_{A},M_{B},e}=\left( \zeta _{+}\left\vert 1_{A}\right\rangle
\left\vert 0_{B}\right\rangle +\eta _{+}\left\vert 0_{A}\right\rangle
\left\vert 1_{B}\right\rangle \right) \left( H.c.\right) +\left(
1-\left\vert \zeta _{+}\right\vert ^{2}-\left\vert \eta _{+}\right\vert
^{2}\right) \left\vert 0_{A}\right\rangle \left\vert 0_{B}\right\rangle
\left\langle 0_{A}\right\vert \left\langle 0_{B}\right\vert ,
\end{equation*}%
where $H.c.$ stands for \emph{Hermitian conjugate} and 
\begin{eqnarray*}
\zeta _{+} &=&\frac{e^{i\phi_1}f\left( \tau\right) +l\left( \tau\right) 
}{\sqrt{2}},\qquad \eta _{+}=\frac{f\left( \tau\right) +e^{i\phi
_{1}}l\left( \tau\right) }{\sqrt{2}}, \\
f\left( t\right)  &=&\frac{e^{-i\omega t}}{2}\left( e^{-\left(
k+k_{c}\right) t}+e^{-\left( k-k_{c}\right) t}\right) ,\qquad l\left(
t\right) =\frac{e^{-i\omega t}}{2}\left( e^{-\left( k+k_{c}\right)
t}-e^{-\left( k-k_{c}\right) t}\right) .
\end{eqnarray*}%
Then, a second atom, initially in the state $g$, absorbs the antisymmetric
field mode by interacting with the mode $M_{A}$ during three $\pi $ Rabi pulses
and with the mode $M_{B}$ for one $\pi /2$ Rabi pulse. With respect to $M_{A}$,
the interaction time can be adjusted by removing the atomic transition from
the resonance with this mode, by means of the Stark effect, in the beginning of
the path of the atom inside the cavity $A$, and then by letting in the resonance
for the time necessary for three $\pi $ Rabi pulses. As regards $M_{B}$, one
can allow the atom to interact with this mode during the time required for the 
$\pi /2$ Rabi pulse, in the beginning of its path inside the cavity $B$,
after which the interaction is interrupted using the Stark effect. When the
atom leaves the cavity $B$, $M_{A}$ is in the vacuum state and the second atom
plus $M_{B}$ state can be written as 
\begin{eqnarray*}
\rho _{M_{B},A_{2},e} &=&\frac{1}{2}\left\{ \left[ e^{-i\phi_2}\left(
\zeta _{+}-\eta _{+}\right) \left\vert e_{2}\right\rangle \left\vert
0_{B}\right\rangle +\left( \zeta _{+}+\eta _{+}\right) \left\vert
g_{2}\right\rangle \left\vert 1_{B}\right\rangle \right] \left[ H.c.\right]
\right.  \\
&&\left. +2\left( 1-\left\vert \zeta _{+}\right\vert ^{2}-\left\vert \eta
_{+}\right\vert ^{2}\right) \left\vert g_{2}\right\rangle \left\vert
0_{B}\right\rangle \left\langle g_{2}\right\vert \left\langle
0_{B}\right\vert \right\} ,
\end{eqnarray*}%
where $e^{-i\phi_2}$ corresponds to the phase accumulation during the
Stark effect in cavity $B$. 
If we consider that the first
atom was measured in the state $g$, we reach a final state concerning the
second atom and the mode $M_{B}$ given by 
\begin{eqnarray*}
\rho _{M_{B},A_{2},g} &=&\frac{1}{2}\left\{ \left[ e^{-i\phi_2}\left(
\zeta _{-}-\eta _{-}\right) \left\vert e_{2}\right\rangle \left\vert
0_{B}\right\rangle +\left( \zeta _{-}+\eta _{-}\right) \left\vert
g_{2}\right\rangle \left\vert 1_{B}\right\rangle \right] \left[ H.c.\right]
\right.  \\
&&\left. +2\left( 1-\left\vert \zeta _{-}\right\vert ^{2}-\left\vert \eta
_{-}\right\vert ^{2}\right) \left\vert g_{2}\right\rangle \left\vert
0_{B}\right\rangle \left\langle g_{2}\right\vert \left\langle
0_{B}\right\vert \right\} ,
\end{eqnarray*}%
where 
\begin{equation*}
\zeta _{-}=\frac{-e^{i\phi_1}f\left( \tau\right) +l\left( \tau\right) }{%
\sqrt{2}},\qquad 
\eta _{-} =\frac{f\left( \tau\right)
-e^{i\phi_1}l\left( \tau\right) }{\sqrt{2}}.
\end{equation*}%
Therefore, the probabilities of measuring the first atom in the state $x$
and the second one in the state $y$ (where $x$ and $y$ stand for $e$ or $g$%
) are 
\begin{eqnarray}
P_{ee} &=&\frac{1}{4}\left( 1-\cos \phi_1\right) e^{-2\left(
k-k_{c}\right) \tau},  \label{Prob1} \\
P_{eg} &=&\frac{1}{2}-P_{ee},  \notag \\
P_{ge} &=&\frac{1}{4}\left( 1+\cos \phi_1\right) e^{-2\left(
k-k_{c}\right) \tau},  \notag \\
P_{gg} &=&\frac{1}{2}-P_{ge}.  \notag
\end{eqnarray}

In equations (\ref{Prob1}) we see that the interference is completely
recovered in two cases. One of them corresponds to $\tau=0$, i.e., there is no
interaction with the environment. The other is the limiting case $k_{c}=k$,
where the environment does not disturb the erasure process, since it
interacts only with the symmetric mode \cite{art25} and the erasure is based
on the absorption of the antisymmetric mode.

\section{Investigating the cross decay rates}

In the sequence described above, the second atom absorbs energy only from
the antisymmetric mode. If we change this scheme so that the second atom
interacts with the mode $M_{A}$ during the time of one $\pi $ Rabi pulse,
and maintain the remainder unmodified, it will absorb energy from the symmetric
mode only. In this case, the limiting case $k_{c}=k$ leads to the maximum
attenuation of the interference fringes. Indeed, the probabilities of
measuring the state of both atoms, defined analogously as in equation (\ref%
{Prob1}), are 
\begin{eqnarray}
P_{ee}^{\prime } &=&\frac{1}{4}\left( 1+\cos \phi_1\right) e^{-2\left(
k+k_{c}\right) \tau},  \label{Prob2} \\
P_{eg}^{\prime } &=&\frac{1}{2}-P_{ee}^{\prime },  \notag \\
P_{ge}^{\prime } &=&\frac{1}{4}\left( 1-\cos \phi_1\right) e^{-2\left(
k+k_{c}\right) \tau},  \notag \\
P_{gg}^{\prime } &=&\frac{1}{2}-P_{ge}^{\prime }.  \notag
\end{eqnarray}%
The interference decreases
according to $e^{-2\left( k-k_{c}\right) \tau}$ in equations (\ref{Prob1}), and according to $e^{-2\left( k+k_{c}\right) \tau}$ in equations (\ref%
{Prob2}); this result is
related to the fact that the environment acts more
strongly on the symmetric mode than on the antisymmetric mode \cite{art25}. This can be used to
measure the cross-decay rate $k_{c}$. Once the two experimental schemes have
been completed, the measures of the frequency of the atomic states can be aggregated
by computing the quantity 
\begin{eqnarray*}
\xi  &=&P_{ge}-P_{gg}-P_{ee}+P_{eg}+P_{ge}^{\prime }-P_{gg}^{\prime
}-P_{ee}^{\prime }+P_{eg}^{\prime } \\
&=&e^{-2\left( k-k_{c}\right) \tau}\left( 1-e^{-4k_{c}\tau}\right) \cos
\phi_1,
\end{eqnarray*}%
which will be non-zero if, and only if, $k_{c}$ is not null.

\section{Conclusion}

We explored the effects of the environment on quantum erasure in the cavity
quantum electrodynamics domain. We showed that the bath disturbs the erasure
process in a way that may depend on the details of the experimental setup.
In fact, the attenuation of the interference fringes due to the environment
depends, for non zero cross decay rates, on the mode (symmetric or
antisymmetric) absorbed by the eraser, namely, the second atom that crosses
the apparatus. This can be used to estimate the cross decay rates, which are
related to microscopic correlations in the system-environment interaction.
These rates, whose conditions for existence may be associated with the
construction of modes spatially close in the scale of their wavelengths, are
responsible for superradiance and subradiance and, in the limit, for
decoherence-free subspaces. In this limit, the antisymmetric mode decouples
from the bath, and an erasure scheme not affected by the environment can be
envisaged.

\section{Acknowledgements}

ARBM and JGPF acknowledge the support from the 
Brazilian agencies CNPq (Grants 486920/2012-7, 
305380/2012-5 and 306871/2012-2) and CEFET/MG (PROPESQ program---Grant 10122\_2012). 
RRJr acknowledges the Brazilian agency 
FAPEMIG (Grant APQ-00597-12) for partial financial support.

\end{document}